\documentclass[11pt, reqno]{amsart}
\usepackage{geometry}
\usepackage{graphicx}
\usepackage{amssymb}
\usepackage{amsmath, mathrsfs, stmaryrd}
\usepackage{comment, tensor}
\usepackage{color}
\newtheorem{theorem}{Theorem}
\newtheorem{definition}[theorem]{Definition}

\newcommand{\R}{\mathbb R}

\newcommand{\na}{\nabla}
\newcommand{\lt}{\left}
\newcommand{\rt}{\right}
\newcommand{\cb}{\underline{c}}

\newcommand{\rw}{\rightarrow}
\usepackage{xpatch}
\usepackage{hyperref}
\usepackage{footnotebackref}

\makeatletter
\xpretocmd{\@adminfootnotes}{\let\@makefntext\BHFN@OldMakefntext}{}{}
\renewcommand\@makefntext[1]{%
  \@ifundefined{@makefnmark}
    {}
    {%
     \renewcommand\@makefnmark{%
       \mbox{%
         \textsuperscript{%
           \normalfont
           \hyperref[\BackrefFootnoteTag]{\@thefnmark}%
         }%
       }\,%
     }%
     \BHFN@OldMakefntext{#1}%
  }%
}
\makeatother

\numberwithin{theorem}{section}
\numberwithin{equation}{section}

\title[conserved quantities at infinity]{Conserved quantities in general relativity- the view from null infinity}

\author{Po-Ning Chen, Mu-Tao Wang, Ye-Kai Wang, and Shing-Tung Yau}

\begin{document}
\begin{abstract}

In general relativity, an idealized distant observer is situated at future null infinity where light rays emitted from the source approach. 
This article concerns conserved quantities such as mass, energy-momentum, angular momentum, and center of mass at future null infinity. The classical definitions of Bondi mass at future null infinity ascertains the mass radiated away in gravitational waves distinctively. However, the same question for other conserved quantities such as angular momentum has been a subtle issue since the discovery of ``supertranslation ambiguity" in the 1960's. 
Recently, new definitions of angular momentum and center of mass were proposed and proved to be free of such ambiguity \cite{CWWY1, CKWWY}.
These new definitions arise as limits of the Chen-Wang-Yau quasilocal conserved quantities, which are based on the theory of optimal isometric embedding and quasilocal mass of Wang-Yau. It is the purpose of this note to discuss these recent developments.\end{abstract}

\thanks{P.-N. Chen is supported by Simons Foundation collaboration grant \#584785, M.-T. Wang is supported by NSF grant DMS-1810856 and DMS 2104212, Y.-K. Wang is supported by Taiwan MOST grant 109-2628-M-006-001-MY3, and S.-T. Yau is supported by the John Templeton Foundation (award number 61497). This material is based upon work supported by the National Science
Foundation under Grant Number DMS 1810856 and DMS 2104212 (Mu-Tao\ Wang).  }\maketitle
\section{Introduction}

In the theory of general relativity, the definition of conserved quantities has been a challenging task.  An essential difficulty goes back to Einstein's equivalence principle, which asserts that gravitation, unlike other physical fields, cannot be localized \cite[\S 20.4]{MTW}. Moreover the issue is complicated by the fact that general relativity is a nonlinear theory and there is no canonical choice of a reference system. An isolated gravitating system corresponds to an asymptotically flat spacetime where gravitation is weak at infinity. In terms of an asymptotically flat coordinate system, there are well-defined notions of energy/mass, most notably the ADM energy/mass at spatial infinity and the Bondi energy/mass at null infinity. However, defining other conserved quantities such as angular momentum and center of mass is more subtle. At future null infinity, ``supertranslation ambiguity" has presented an essential difficulty for these attempts since 1960's.  According to Penrose \cite{Penrose2}, the very concept of angular momentum gets shifted by these supertranslations and ``it is hard to see in these circumstances how one can rigorously discuss such questions as the angular momentum carried away by gravitational radiation" (page 654 of \cite{Penrose2}). Recently the authors \cite{CWWY1, CKWWY} proposed the first definitions of angular momentum and center of mass at future null infinity that are completely free from supertranslation ambiguity. 
Our approach of defining these conserved quantities at future null infinity is to take the limits of quasilocal definitions. In this article, we first review the concepts of conserved quantities in special relativity and general relativity and then discuss the theory of quasilocal mass and optimal isometric embedding proposed by Wang-Yau in \cite{Wang-Yau1,Wang-Yau2} and the definition of Chen-Wang-Yau quasilocal conserved quantities in \cite{CWY3, CWY4}. At the end, we explain how the limits of the Chen-Wang-Yau quasilocal conserved quantities at null infinity provide viable definitions of conserved quantities free of any supertranslation ambiguity.

\section{Conserved quantities in relativity}

Relativity is a unified theory of space and time. The spacetime of special relativity is the Minkowski spacetime $\mathbb{R}^{3,1}=\mathbb{R}\times \mathbb{R}^3$ with Lorentz metric of signature $(-1, 1, 1, 1)$. We normalize  the speed of light to be $1$. The light cone consists of four-vectors $(t, x^i), i=1, 2, 3$ with $-t^2+(x^1)^2+(x^2)^2+(x^3)^2=0$. As nothing travels faster than light, a material particle or an observer moves in future timelike direction.

To each matter field, an energy-momentum tensor $T$ is attached. $T$ is derived from the Lagrangian of the field and is described by first derivatives of the field. In particular, it is a symmetric $(0,2)$ tensor $T_{\mu\nu}$ which satisfies the conservation law

\begin{equation}\label{conserv}\nabla^\mu T_{\mu\nu}=0.\end{equation}
The conserved quantities of a physical system $\Omega$ in the Minkowski spacetime are obtained by integrating $T$ on $\Omega$ with respect to Killing fields. To be more precise, given a spacelike bounded region $\Omega$, the conserved quantity of $\Omega$ with respect to a Killing field $t^\nu$ is the flux integral
\[\int_\Omega T_{\mu\nu} t^\mu u^\nu \] where $u^\nu$ is the future timelike unit normal of $\Omega$. 

In general, one defines conserved quantities with respect to a Killing field in the 10 dimensional Poincar\'e algebra of the Minkowski spacetime, which correspond to the infinitesimal symmetry of the Poincar\'e group. We recall that a translation Killing field is a linear combination of $\frac{\partial}{\partial t}, \frac{\partial}{\partial x^i}, i=1, 2, 3$. 
A rotation Killing field is the image of $x^i\frac{\partial}{\partial x^j}-x^j\frac{\partial}{\partial x^i}, i<j$ 
under a Lorentz transformation and a boost Killing field is the image of $x^i\frac{\partial}{\partial t}+t \frac{\partial}{\partial x^i}, i=1, 2, 3$ under a Lorentz transformation. These Killing fields define energy-momentum, angular momentum, and center of mass.

In general relativity, spacetime is a 4-dimensional manifold with a Lorentz metric $g$, the gravitational field. The local causal structure of spacetime remains the same, and each tangent space is isometric to the Minkowski space. The relation between the gravitation field and matter fields is exactly described by the Einstein equation

    \begin{equation}\label{einstein}Ric-\frac{1}{2} R g=8\pi T\end{equation}
     where $Ric$ is the Ricci curvature, and $R$ is the scalar curvature of $g$, respectively. $T$ represents the energy-momentum tensor of all matter fields and satisfies the conservation law \eqref{conserv} with respect to the Lorentz metric $g$. 

Concerning energy, one seeks an energy momentum tensor for gravitation. However, it turns out that the first derivatives of $g$ are all coordinate dependent, and thus there is no density for gravitational energy. This is Einstein's equivalence principle.
One can still integrate $T$ on the right hand side of \eqref{einstein} but this gives only the energy contribution from matter. Indeed, there exist vacuum spacetimes, i.e. $T=0$, with nonzero energy such as Schwarzschild's or Kerr's solution of Einstein's equation. This is gravitational energy by the sheer presence of spacetime curvature. Even without energy density, one can still ask the question: what is the energy in a system $\Omega$, counting contribution from gravitation and all matter fields?

 In special relativity, the energy integral of $T$ on $\Omega$ depends only on the boundary data by energy conservation. One expects energy conservation in general relativity as well, and thus this information should be encoded in the geometry of the two-dimensional boundary surface $\Sigma=\partial\Omega$. This leads to the well-known problem of quasilocal energy/mass in general relativity. The first one in Penrose's 1982 list \cite{Penrose2} of major unsolved problems in classical
general relativity is ``Find a suitable quasilocal definition of energy-momentum in general relativity". In the next section, we review the theory of Wang-Yau quasilocal mass on which the definitions of the Chen-Wang-Yau quasilocal conserved quantities are based.

\section{Wang-Yau quasilocal mass and Chen-Wang-Yau quasilocal conserved quantities}
 The notion of quasilocal mass is attached to a 2-dimensional closed surface $\Sigma$ which bounds a spacelike region in spacetime. $\Sigma$ is assumed to be a topological 2-sphere, but with different intrinsic geometry and extrinsic geometry, we expect to read off the effect of gravitation in the spacetime vicinity of the surface. Suppose the surface is spacelike, i.e. the induced metric $\sigma_\Sigma$ is Riemannian.  An essential part of the extrinsic geometry is measured by the mean curvature vector field $\bf{H}$ of $\Sigma$. $\bf{H}$ is a normal vector field of the surface such that the null expansion along any null normal direction $\ell$ is given by the pairing $\langle {\bf H}, \ell\rangle$ of
$\bf{H}$ and $\ell$.

In \cite{Wang-Yau1}, Wang-Yau proposed the following definition of quasilocal mass which depends only on $\sigma_\Sigma$ and $\bf{H}$ of a 2-surface $\Sigma$ in spacetime.  To evaluate the quasilocal mass of $\Sigma$ with the physical data $(\sigma_\Sigma, \bf{H})$, one first solves the optimal isometric embedding equation, see \eqref{oiee} below, which gives an embedding of $\Sigma$ into the Minkowski spacetime with the image
surface $\Sigma_0$ that has the same 
induced metric as $\Sigma$, i.e. $\sigma_\Sigma$. One then compares the extrinsic geometries of $\Sigma$ and $\Sigma_0$ and evaluates the quasilocal mass from $\sigma_\Sigma, \bf{H}$ and $\bf{H_0}$.

Assuming the mean curvature vector ${\bf H}$ is spacelike, the physical surface $\Sigma$ with physical data $(\sigma_\Sigma, \bf{H})$ gives $(\sigma_\Sigma, |\bf{H}|, \alpha_{\bf {H}})$ where $|{\bf H}|>0$ is the Lorentz norm of $\bf{H}$ and $\alpha_{\bf H}$ is the connection one-form determined by $\bf{H}$. Given an isometric embedding $X:\Sigma\rightarrow \R^{3,1}$ of $\sigma_\Sigma$, let $\Sigma_0$ be the image $X(\Sigma)$ and $(\sigma_\Sigma, |\bf{H}_0|, \alpha_{\bf {H}_0})$ be the corresponding data of $\Sigma_0$ (${\bf H_0}$ is again assumed to be spacelike).

Let $T$ be a future timelike unit Killing field of $\R^{3,1}$ and define $\tau=-\langle X, T\rangle$ as a function on $\Sigma$. Define a function $\rho$ and a 1-form $j_a$ on $\Sigma$:
  \[ \begin{split}\rho &= \frac{\sqrt{|{\bf H}_0|^2 +\frac{(\Delta \tau)^2}{1+ |\nabla \tau|^2}} - \sqrt{|{\bf H}|^2 +\frac{(\Delta \tau)^2}{1+ |\nabla \tau|^2}} }{ \sqrt{1+ |\nabla \tau|^2}}\\
 j_a&=\rho {\nabla_a \tau }- \nabla_a \left( \sinh^{-1} (\frac{\rho\Delta \tau }{|{\bf H}_0||{\bf H}|})\right)-(\alpha_{{\bf H}_0})_a + (\alpha_{{\bf H}})_a, \end{split}\] where $\nabla_a$ is the covariant derivative with respect to the metric $\sigma_\Sigma$, $|\nabla \tau|^2=\nabla^a \tau\nabla_a \tau$ and $\Delta \tau=\nabla^a\nabla_a \tau$. 
$\rho$ is the quasilocal mass density and $j_a$ is the quasilocal momentum density. A full set of quasilocal conserved quantities was defined in \cite{CWY3, CWY4} using $\rho$ and $j_a$. 
 
The optimal isometric embedding equation for $(X, T)$ is 
\begin{equation}\label{oiee} \begin{cases}
\langle dX, dX\rangle&=\sigma_\Sigma \\
\nabla^a j_a&=0.
\end{cases}\end{equation}
The first equation is the isometric embedding equation into the Minkowski spacetime and the second one is the Euler-Lagrange equation of the quasilocal energy $E(\Sigma, \tau)$ \cite{Wang-Yau1,Wang-Yau2} in the space of isometric embeddings.
The quasi-local mass for the optimal isometric embedding $(X, T)$ is defined to be \[E(\Sigma, X, T)=\frac{1}{8\pi}\int_\Sigma \rho.\]
 It is shown in \cite{Wang-Yau1, Wang-Yau2} that $E(\Sigma, X, T)$ is { positive in general}, and { zero for surfaces in the Minkowski spacetime}. 

The theory of quasilocal mass and optimal isometric embedding was employed by Chen-Wang-Yau in \cite{CWY3, CWY4} to define quasilocal conserved quantities. For an optimal isometric embedding $(X, T)$, by restricting a rotation (or boost) Killing field $K$ of $\R^{3,1}$ to $\Sigma_0=X(\Sigma)\subset \R^{3,1}$, the quasi-local conserved quantity is defined to be: 
\[-\frac{1}{8\pi} \int_\Sigma \langle K, T\rangle \rho+(K^\top)^a  j_a ,\] where $K^\top$ is the component of $K$ that is tangential to $\Sigma_0$.
In particular, $K=x^i\frac{\partial}{\partial x^j}-x^j\frac{\partial}{\partial x^j}, i<j$ defines an angular momentum with respect to $\frac{\partial}{\partial t}$. Here $(t, x^i)$ and $(\frac{\partial}{\partial t}, \frac{\partial}{\partial x^i})$ are standard coordinates and coordinate vectors of the Minkowski spacetime.

The image of the optimal isometric embedding $\Sigma_0$ is essentially the ``unique" surface in the Minkowski spacetime that best matches the physical surface $\Sigma$. 
If the original surface $\Sigma$ happens to be a surface in the Minkowski spacetime, the above procedure identifies $\Sigma_0=\Sigma$ up to a global isometry. 

A solution of the optimal isometric embedding equation is indeed a critical point of the quasilocal energy $E(\Sigma, \tau)$. In \cite{CWY2}, we study the minimizing and uniqueness property for a solution of the optimal isometric embedding equation. In particular, the following theorems hold true:

\begin{theorem}\cite{CWY2}
Let $(\sigma_\Sigma, H)$ be the data of a spacelike surface $\Sigma$ with spacelike mean curvature vector $H$ in the Minkowski spacetime and $T$ be a unit timelike Killing field. Suppose the projection of $\Sigma$ onto the orthogonal complement of $T$ is a convex surface. Then

(1) the kernel of the linearized optimal isometric 
embedding system consists precisely of Lorentz transformations. 

(2) the second variation of the quasilocal energy  $E(\Sigma, \tau)$
is non-negative definite. \end{theorem}

For a spacelike surface with spacelike mean curvature vector in a general spacetime, one has

 \begin{theorem} \cite{CWY2} Let $(\sigma_\Sigma, H)$ be the data of a spacelike surface $\Sigma$ in a general spacetime. Suppose that $\tau_0$ is a critical point of the quasi-local energy $E(\Sigma, \tau)$ and that the corresponding quasilocal mass density $\rho$ is positive, then $\tau_0$ is a local minimum for  $E(\Sigma,\tau)$.
\end{theorem}

The above theorems allow us to solve the optimal isometric embedding system for configurations that limit
to a surface in the Minkowski spacetime. This is in particular sufficient for calculations at infinity of an isolated system.

\section{Conserved quantities at null infinity}
In this section, we focus on the definition of total conserved quantities at future null infinity $\mathscr{I}^+$. We will first review the well-known description of null infinity by the Bondi-Sachs coordinate system, in terms of which the supertranslation ambiguity was first discovered. Such ambiguity is indeed ubiquitous in any description of the future null infinity.  We then discuss the supertranslation invariance property of the Bondi-Sachs energy-momentum. At the end, we discuss how the limit of the CWY quasilocal conserved quantities gives the definition of conserved quantities at null infinity that are free of supertranslation ambiguity.

\subsection{The description of null infinity}
 
The spacetime near $\mathscr{I}^+$ can be described in terms of the Bondi-Sachs coordinates $(u, r, \theta, \phi)$ which are chosen in the following way.  Level sets of $u$ are null hypersurfaces generated by null geodesics, $\theta$ and $\phi$ are extended  by constancy along the integral curves of the gradient vector field of $u$, and 
 $r$ is chosen according to the following determinant condition. 
In terms of a Bondi-Sachs coordinate system $(u, r,  x^2=\theta, x^3=\phi)$, the spacetime metric takes the form
\begin{equation}\label{spacetime_metric}g_{\alpha\beta}dx^\alpha dx^\beta= -UV du^2-2U dudr+r^2 h_{ab}(dx^a+W^a du)(dx^b+W^b du).\end{equation} The index conventions here are $\alpha, \beta=0,1, 2, 3$, $a, b=2, 3$, and $u=x^0, r=x^1, \theta=x^2, \phi=x^3$. See \cite{MW} for more details of the construction of the coordinate system. 
The metric coefficients $U, V, h_{ab}, W^a$  of \eqref{spacetime_metric} depend on $u, r, \theta, \phi$, and the determinant condition implies that $\det h_{ab}$ is independent of $u$ and $r$. These gauge conditions reduce the number of metric coefficients of a Bondi-Sachs coordinate system to six (there are only two independent components in $h_{ab}$). On the other hand, the boundary conditions $U\rightarrow 1$, $V\rightarrow 1$, $W^a\rightarrow 0$, $h_{ab}\rightarrow \sigma_{ab}$ are imposed as $r\rightarrow \infty$, where $\sigma_{ab} = d\theta^2 +\sin^2\theta \,d\phi^2$ stands for the standard metric on $S^2$. The special gauge choice of the Bondi-Sachs coordinates implies a hierarchy among the vacuum Einstein equations, see \cite{MW, HPS}.

 Assuming the outgoing radiation condition \cite{Sachs, VDB, MW, Winicour, VK}, the boundary condition and the vacuum Einstein equation imply that as $r\rightarrow \infty$, all metric coefficients can be expanded in inverse integral powers of $r$.\footnote{The outgoing radiation condition assumes the traceless part of the $r^{-2}$ term in the expansion of $h_{ab}$ is zero. The presence of this traceless term will lead to a logarithmic term in the expansions of $W^a$ and $V$. Spacetimes with metrics which admit an expansion in terms of $r^{-j}\log^i r$ are called ``polyhomogeneous" and are studied in \cite{CMS}. They do not obey the outgoing radiation condition or the peeling theorem \cite{VK}, but they do appear as perturbations of the Minkowski spacetime by the work of Christodoulou-Klainerman \cite{CK}.} In particular, 
\[\begin{split} U&=1+O(r^{-2}),\\
V&=1-\frac{2m}{r}+O(r^{-2}),\\
W^a&=O(r^{-2}),\\
h_{ab}&={\sigma}_{ab}+\frac{C_{ab}}{r}+O(r^{-2}),\end{split}\] where  $m=m(u, x^a)$ is the mass aspect and $C_{ab}=C_{ab}(u, x^a)$ is the shear of this Bondi-Sachs coordinate system. In addition, $N_{ab}=\partial_u C_{ab}$ is called the news. 
The Bondi-Sachs  energy-momentum 4-vector associated with a $u$-slice is then \[E(u)=\frac{1}{4\pi}\int_{S^2_\infty} m(u, x^a) dv_\sigma,\,\, P^i(u)=\frac{1}{4\pi}\int_{S^2_\infty} m(u, x^a) \tilde{X}^i dv_\sigma, i=1,2, 3\] where $\{\tilde{X}^i=\tilde{X}^i(x^a), i=1, 2, 3\}$ is an orthonormal basis of the $(-2)$ eigenspace of $\Delta=\Delta_\sigma$ (these are the usual $\ell=1$ spherical harmonics) and $dv_\sigma$ is the area form of the metric $\sigma$. The positivity of the Bondi mass

 \begin{equation}\label{pmt} \sqrt{E^2-\sum_i (P^i)^2}\end{equation} was proved by Schoen-Yau \cite{SY} and Horowitz-Perry \cite{HP} under the dominant energy condition and a global assumption on the horizon, see also \cite{CJS}. The Einstein equations imply the following equation satisfied by the mass aspect along $\mathscr{I}^+$:
\begin{equation}\label{du_mass_aspect}\partial_u m=\frac{1}{4}\nabla^a\nabla^b (N_{ab})-\frac{1}{8}|N_{ab}|_\sigma^2,\end{equation} where $|N_{ab}|_\sigma^2=\sigma^{ac}\sigma^{bd} N_{ab}N_{cd}$. 
Integrating over $S^2_\infty$ with the metric $\sigma$ yields the well-known Bondi mass loss formula \cite{Bondi}:
\begin{equation}\label{energy_loss1}\frac{d}{du} E(u)=-\frac{1}{32\pi}\int_{S^2_\infty} |N_{ab}|_\sigma^2 dv_\sigma \leq 0.\end{equation} In particular,
\begin{equation} \label{energy_loss2} E(u_1)\leq E(u_0) \text{ if } u_1\geq u_0.\end{equation}


This formula indeed corresponds to energy loss, see \cite{HYZ} for a monotonicity formula for the quantity $E-\sqrt{\sum_i (P^i)^2}$.

\subsection{Supertranslation invariance of the Bondi mass}
Rescaling the spacetime metric \eqref{spacetime_metric} by $r^{-2}$ as $r\rightarrow \infty$, the limit of $r^{-2}g_{\alpha\beta} dx^\alpha dx^\beta$ approaches  $ \sigma_{ab} dx^a dx^b$, or the null metric on $\mathscr{I}^+$.\footnote{This is a special case of conformal compactification. In general, the metric on the unphysical spacetime is of the form $\Omega^2 g_{\alpha\beta} dx^\alpha dx^\beta$ and $\Omega=0$ corresponds to $\mathscr{I}^+$, see \cite{Penrose3, Penrose4, Geroch}.}  Therefore, $\mathscr{I}^+$ can be viewed as a null three-manifold:  
\[ \mathscr{I}^+= I\times (S^2, \sigma_{ab})\] with $u\in I$, $x^a\in S^2$. 

Each spacetime Bondi-Sachs coordinate system $(u, r, x^a)$ induces such a limiting coordinate system $(u, x^a)$ on $\mathscr{I}^+$, together with the mass aspect $m(u, x^a)$ and the shear $C_{ab}(u, x^a)$. In the following, as long as no confusion arises, we shall not distinguish the Bondi-Sachs coordinate system $(u, r, x^a)$ on the spacetime near $\mathscr{I}^+$ and the induced coordinate system $(u, x^a)$ on $\mathscr{I}^+$, and shall abbreviate them as $(u, r, x)$ and $(u, x)$, respectively. Such a Bondi-Sachs coordinate system is by no means unique and the BMS group, which corresponds to the diffeomorphism group that preserves the gauge and boundary conditions, acts on the set of Bondi-Sachs coordinate systems.

The symmetry of $\mathscr{I}^+$ consists of the Bondi-Metzner-Sachs (BMS) group, of which supertranslations form an infinite dimensional subgroup. 
\begin{definition} Suppose $(\bar{u}, \bar{x})$ and $(u, x)$ are two Bondi-Sachs coordinate systems on $\mathscr{I}^+$. 
A BMS transformation $\mathscr{I}^+ \rw \mathscr{I}^+$ is given by
\begin{align}\label{BMS_transformation}
(u, x) = \lt( K(\bar x) \bar u + f(\bar x), g(\bar x) \rt)
\end{align}
where $f$ is a function on $S^2$ and $g: (S^2, \bar{\sigma}) \rw (S^2, \sigma)$ is a conformal map with $g^*\sigma = K^2 \bar \sigma$.

A BMS transformation is called a {\it supertranslation} if $g$ is the identity map, $K\equiv 1$, and $(u, x)=(\bar{u}+f(\bar{x}), \bar{x})$. 

\end{definition}

The infinitesimal symmetry of $\mathscr{I}^+$ is the BMS algebra of BMS fields, see \cite{CWWY2} for a complete description of BMS fields and their extensions.

A {\it supertranslation} is thus a change of Bondi-Sachs coordinates $(\bar{u}, \bar{x})\rightarrow (u, x)$ such that 
\begin{equation}\label{coord_change} u = \bar u + f (x), x=\bar{x}\end{equation} on $\mathscr{I}^+$ for a function $f$ that is defined on $S^2$.

We assume $\mathscr{I}^+$ extends from $i^0$ ($u=-\infty$) to $i^+$ ($u=+\infty$) and that there exists a constant $\varepsilon>0$ such that
\begin{equation}\label{news_decay}
N_{ab}( u,x) = O(|u|^{-1-\varepsilon}) \mbox{ as } u \rw \pm\infty.
\end{equation}

Therefore, by \eqref{energy_loss1},  the total flux of $E(u)$ is 
\begin{equation}\label{flux} \lim_{u\rightarrow +\infty} E(u)-\lim_{u\rightarrow -\infty} E(u)=-\frac{1}{32\pi}\int_{-\infty}^{+\infty}  \int_{S^2_\infty} |N_{ab}|_\sigma^2 dv_\sigma du  \end{equation}

For a supertranslation \eqref{coord_change}, it is known that the news $ \bar{N}_{ab}(\bar u,x)$  in the $(\bar u, \bar{x})$ coordinate system are related to the news $ {N}_{ab}(u,x)$  in the $(u, x)$ coordinate system through:
\begin{equation}\label {news}\bar{N}_{ab}(\bar u,x) = N_{ab}(\bar u+f(x),x).\end{equation} See \cite[(C.117) and (C.119)]{CJK} for example. Together with \eqref{news_decay} and \eqref{flux}, the supertranslation invariance of the total flux of mass of $E(u)$ follows.

\subsection{Definitions of angular momentum and center of mass at null infinity}

There have been various proposals of angular momentum at future null infinity since the 1960's. Different approaches (Hamiltonian, spinor-twistor, Komar type etc.) have led to different definitions such as Winicour-Tamburino 1965 \cite{WT}, Newmann-Penrose 1966 \cite{NP2}, Bramson 1975 \cite{Bramson}, Ashtekar-Hansen 1978 \cite{AH}, Penrose 1982 \cite{Penrose1},  Ludvigsen-Vickers 1983 \cite{LV}, 
Dray-Streubel 1984 \cite{DS}, Moreschi 1986 \cite{Moreschi}, Dougan-Mason 1991 \cite{DM}, Rizzi 1997 \cite{Rizzi}, Chru\'sciel-Jezierski-Kijowski 2002 \cite{CJK}, Barnich-Troessaert 2011\cite{BT}, Hawking-Perry-Strominger 2017 \cite{HPS}, Klainerman-Szeftel 2019 \cite{KS}, etc. Each one of them recovers the angular momentum of the model rotating spacetime of Kerr, but the issue of supertranslation ambiguity remains unsolved. There were efforts to eliminate supertranslation ambiguity by choosing special
foliations of $\mathscr{I}^+$, for example,  ``nice sections"by Moreschi \cite{Moreschi},  or ``preferred cuts" by Rizzi \cite{Rizzi}. However, they work only in special cases which seem to exclude black hole formation \cite[reference item 14]{Rizzi}.

These conserved quantities are defined with respect to rotation BMS fields (angular momentum) and boost BMS fields (center of mass). In terms of a Bondi-Sachs coordinate system $(u, x^a)$ at future null infinity $\mathscr{I}^+$, 
a rotation BMS field is given by $Y=\epsilon^{ab}\nabla_b \tilde{X}^k \frac{\partial}{\partial x^a}$ and a boost BMS field is given by  $Y=\nabla^a \tilde{X}^k\frac{\partial}{\partial x^a}+u\tilde{X}^k\frac{\partial}{\partial u}$. These are infinitesimal symmetries of the future null infinity $\mathscr{I}^+$. For a complete description of BMS fields, see \cite{CWWY2}.

Regarding the definition of angular momentum and center of mass, one needs the angular momentum aspect $N^a$ which appears in further expansions of $W^a$ in \eqref{spacetime_metric} (we follow the convention in \cite{KWY}):
\[ W^a=\frac{1}{2r^2} \nabla^b C_{ab}+r^{-3} \left(\frac{2}{3}N^a-\frac{1}{16}\nabla^a(C_{de} C^{de})-\frac{1}{2} C_b^{\,\,a} \nabla^d C_d^{\,\, b}\right)+O(r^{-4}).\] We also denote $N_a=\sigma_{ab} N^b$.


The calculation of the limit of the Chen-Wang-Yau quasilocal conserved quantities in Bondi-Sachs coordinates was taken up  by Keller-Wang-Yau in  \cite{KWY}. The expression depends on the Hodge decomposition of $C_{ab}$. Write
\[ C_{ab}=\nabla_a\nabla_b {c}-\frac{1}{2}\sigma_{ab} \Delta {c}+\frac{1}{2}(\epsilon_{ad} \nabla^d \nabla_b \underline{ c}+\epsilon_{bd} \nabla^d \nabla_a \underline{c}).\]

The limit of the quasilocal angular momentum defined in \cite{CWY3, CWY4} gives a new definition of angular momentum at null infinity \cite{KWY, CKWWY}. 
The angular momentum of a $u$ cut and with respect to a rotation BMS field $Y=Y^a  \frac{\partial}{\partial x^a}= \epsilon^{ab}\nabla_b \tilde{X}^k \frac{\partial}{\partial x^a}$ is:

\begin{equation}\label{KWY} J(u, Y)=\frac{1}{8\pi} \int_{S^2} Y^a\left(N_a-{c}{\nabla}_a m  -\frac{1}{4} C_{ab}{\nabla}_d C^{db}\right).\end{equation}
 
Comparing with previous definitions, the new definition contains an important correction term in the integrand $c\nabla_a m$ that has never appeared in any previous definitions.  This term comes from solving the optimal isometric embedding equation \eqref{oiee} in the theory of Wang-Yau quasilocal mass \cite{Wang-Yau1, Wang-Yau2} and provides the reference term  that is critical in the Hamiltonian approach of defining conserved quantities. 
  
 The limit of the quasilocal center of mass integral defined in \cite{CWY3, CWY4} also gives a new definition of center of mass integral at null infinity \cite{KWY, CKWWY}. 
The center of mass integral $C(u, Y)$ of a $u$ cut and with respect to a boost BMS field $Y=\nabla^a \tilde{X}^k\frac{\partial}{\partial x^a}+u\tilde{X}^k\frac{\partial}{\partial u}$ is defined to be:
\begin{align}\label{COM}
\begin{split}
 8\pi C(u, Y)&= \int_{S^2} \nabla^{a}\tilde{X} \lt(N_a - \frac{1}{4} C_{ab} \na_d C^{db} - \frac{1}{16} \na_b\lt( C^{de}C_{de}\rt)\rt) -2u\int_{S^2}(\tilde{X} m)  \\
&+\int_{S^2} c\lt(3 \tilde X m - \na^a \tilde{X}  \na_a m\rt) \\
& +\int_{S^2} \lt( 2 \nabla^{a}\tilde{X} \epsilon_{ab} (\na^b \cb) m      - \frac{1}{16} \tilde X \na_a (\Delta+2)\cb \na^a (\Delta+2)\cb\rt)\\
\end{split}
\end{align}

Assuming the decay condition on the news \eqref{news_decay}, the total fluxes of $J(u, Y)$ and $C(u, Y)$ are well-defined as
 $\lim_{u\rightarrow +\infty} J(u, Y)-\lim_{u\rightarrow -\infty} J(u, Y)$ and  $\lim_{u\rightarrow +\infty} C(u, Y)-\lim_{u\rightarrow -\infty} C(u, Y)$, respectively.

All previous definitions of angular momentum on $\mathscr{I}^+$ depend on a specific gauge (a null frame or a spacetime coordinate system). In contrast, the Chen-Wang-Yau definition is geometric and coordinate independent (it depends only on $\sigma, {\bf H}$). In addition, solving the optimal isometric equation is a canonical procedure that is free from any ad hoc referencing or normalization. This leads to the following supertranslation invariance theorem that is proved in \cite{CWWY1, CKWWY}:

\begin{theorem}\label{total_flux_CWY_angular}
Suppose the news tensor decays as
\[ N_{ab}( u,x) = O(|u|^{-1-\varepsilon}) \mbox{ as } u \rw \pm\infty.
\]
Then the total fluxes of $ J(u, Y)$ and $C(u, Y) $ are supertranslation invariant.
\end{theorem}

\subsection{Summary}

Fixing $\tilde{X}^k, k=1, 2, 3$ and denoting $J^k(u)$ by $J(u, Y)$ for $Y=\epsilon^{ab}\nabla_b \tilde{X}^k \frac{\partial}{\partial x^a}$ and $C^k(u)$ by $C(u, Y)$ for $Y=\nabla^a \tilde{X}^k\frac{\partial}{\partial x^a}+u\tilde{X}^k\frac{\partial}{\partial u}  $, we obtain the new definitions of angular momentum $J^k(u)$ and center of mass integral $C^k(u)$. They complement the classical Bondi-Sachs energy momentum $E(u), P^k(u)$ and form a set of conserved quantities that correspond to the Poincar\'e symmetry. All of them can be derived from the limits of quasilocal conserved quantities defined in \cite{Wang-Yau1, Wang-Yau2, CWY3}. The Poincar\'e symmetry is due to the choice of Minkowski reference and is acquired through the reference embedding into the Minkowski spacetime \cite{Wang-Yau1, Wang-Yau2}.

We obtain a complete set of ten conserved quantities $(E, P^k, J^k, C^k)$ at null infinity (all as functions of the retarded time $u$) that satisfy the following properties:

(1) $(E, P^k, J^k, C^k)$ all vanish for any Bondi-Sachs coordinate system of the Minkowski spacetime. 

(2) In a Bondi-Sachs coordinate system of the Kerr spacetime, $P^k$ and $C^k$ vanish, and $E$ and $J^k$ recover the mass and angular momentum. $(E, P^k, J^k, C^k)$ are supertranslation invariant. 

(3) If a spacetime admits a Bondi-Sachs coordinate system such that the news vanishes, then $(E, P^k, J^k, C^k)$ are constant (independent of the retarded time $u$) and supertranslation  invariant. 

(4) On a general spacetime,  the total fluxes of $(E, P^k, J^k, C^k)$ are supertranslation invariant. 

(5) $(E, P^k, J^k, C^k)$ and their fluxes transform according to basic physical laws under ordinary translations (see equations (17) (22) (25) in \cite{CWWY1}).

(6) All of them are expressed in terms of the harmonic mode decompositions and numerical calculations can be readily implemented.

\end{document}